# Process nano scale mechanical properties measurement of thin metal films using a novel paddle cantilever test structure

Chi-Jia Tong & Ming-Tzer Lin

Institute of Precision Engineering,
National Chung Hsing University,
Taichung, 402, Taiwan

*Abstract-*A new technique was developed for studying the mechanical behavior of nano-scale thin metal films on substrate is presented. The test structure was designed on a novel "paddle" cantilever beam specimens with dimensions as few hundred nanometers to less than 10 nanometers. This beam is in triangle shape in order to provide uniform plane strain distribution. Standard clean room processing was used to prepare the paddle sample. The experiment can be operated by using the electrostatic deflection on the "paddle" uniform distributed stress cantilever beam and then measure the deposited thin metal film materials on top of it. A capacitance technique was used to measurement on the other side of the deflected plate to measure its deflection with respect to the force. The measured strain was converted through the capacitance measurement for the deflection of the cantilever. System performance on the residual stress measurement of thin films are calculated with three different forces on the "paddle" cantilever beam, including the force due to the film, compliance force and electrostatic force.

## I. Introduction

Microelectromechanical systems (MEMS) technologies are developing rapidly with increasing study of the design, fabrication and commercialization of microscale systems and devices. The roadmap for MEMS plans for developing the complexity and packing density of devices into the future allowing ever smaller in the range of nanometer scale and more densely packed structures to be fabricated. Continued growth of Microsystems technologies requires still further miniaturization, with a corresponding need to understand how length scales affect mechanical behavior of all the components. Accurate knowledge on the mechanical behaviors of thin film materials used for MEMS is important for successful design and development of MEMS.

Although many previous studies had performed on the characterization of mechanical behavior on thin films, the results obtained from different measurement techniques were vary widely for nominally identical samples due to the difficulty with the techniques. Example such as the nanoindentation experiment on thin film is stronger correlated its substrate. Therefore, the ISO (International Standard Organization)-14577 had defined the proper measurement range of using Nano Indentation test that the contact depth has to be <200nm &

h<t/10, where t is film thickness. As a result, it limited the ability to do extensive study of thin film.

Here a new technique was developed for studying the mechanical behavior of Nano-scale thin metal films on substrate. The test structure was designed on a novel "paddle" cantilever beam specimens with dimensions as few hundred nanometers to less than 10 nanometers. This beam is in triangle shape in order to provide uniform plane strain distribution. The experiment is designed to be operated by using the electrostatic deflection on the "paddle" uniform distributed stress cantilever beam and then measure the deposited thin metal film materials on top of it. Thus allow us to explore mechanical function mechanisms in thin films and film of the thickness on nanoscale regime.

## II. Design & development procedures

In traditional microbeam bending, cantilevered beams of the film of interest are fabricated by micromachining. The end of the beam is deflected using an indentation apparatus which senses vertical displacement. Elastic properties of the thin film material can be determined from the measured load-displacement characteristics [1-3]. If a thin film is deposited on a thicker free-standing cantilever, (e.g. silicon) plastic properties of the thin film can be probed as well [3].

However, the primary drawbacks in common beam bending, due to the non-uniform distribution of stresses, only a small portion of the sample (at the root of the beam) is exposed to the maximum load. One would thus expect to see higher values for yield strength from beam bending experiments compared to techniques which sample larger volumes. In addition, using indentation apparatus providing vertical displacement for the beam will generate localized effects in the vicinity of the loading point due to its destructive force and complicated analysis processes.

The "paddle" cantilever beam approaches here was to design a "constant stress" cantilever beam that eliminate the non-uniform distribution of stresses along the cantilever. In a cantilever beam with a single localized load at the free end, the bending moment varies linearly from zero at the point of load application to a maximum at the built in end. Thus for a parallel sided beam, as in the preceding sketch Fig. 1, the axial stress on the beam surface is proportional to the bending moment; and, is expressed by:

$$\sigma(x) = \frac{M(x) \cdot c}{I} = \frac{6PX}{bt^2} = \frac{PX}{Z}$$

Where: $\sigma(x)$ = bending stress on the beam surface at a distance X from the point of load application, psi (N/m²)
M(x) = bending moment at distance X, in-lbs. (mN)
c = t/2 = distance from neutral axis to beam surface, in(m)
I = bt³/12 = Moment of inertial of beam cross-section, in⁴ (m⁴)
Z = bt²/6 = Section modulus of beam, in³ (m³)
b = beam width , in (m)
t = beam thickness, in (m)

Thus, through the design of a cantilever section which modulus can be made proportional to X by making the width proportional to X, and holding the thickness constant, thus stress is constant from one end of the beam to the other. This constant-stress "paddle" cantilever beam is shown in Figure 1 and was primary discussed in various studies in bulk structure [4].

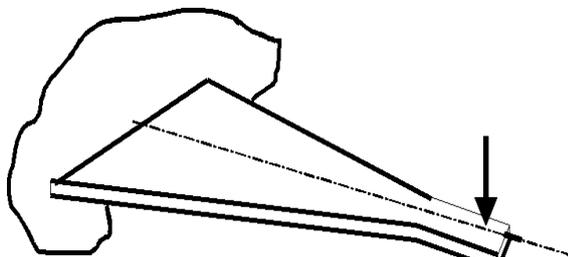

Figure 1: A constant-stress cantilever beam [4]

In addition, the electrostatic deflection force on the designated plate to drag "paddle" cantilever beam can be used instead of using indentation apparatus to provide vertical displacement for the beam. Thus can reduce localized effects in the vicinity of the loading point.

### A. Sample design and fabrication

The single test chip is designed (Figure 2) to fit in an apparatus where one side of the paddle plate is pulled by an electrostatic field. The measured strain was converted through the capacitance measurement for the deflection of the cantilever.

The sample is shown as the structure of a 20mm✕20mm chip with paddle as shown above in plane view and cross section. The thin silicon beam (40 um thick) has a triangular shape in plane view to provide uniform strain in a metallic film on its surface when the beam is bent with force on the paddle. Figure 2 shows the front side and backside view of the sample.

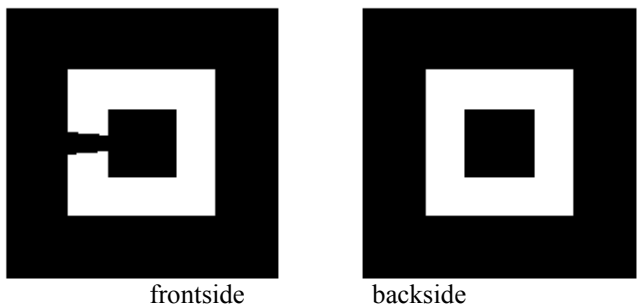

frontside        backside
Figure 2: Schematics view of sample

The sample fabrication procedure is using standard clean room processing for MEMS structures. We utilize these techniques and develop one of the simplest sample fabrication processes to conduct an experiment that can maintain consistent sample preparation and excellent yield procedures. The sample is fabricated on a standard 4 inches diameter double polished silicon wafer. Each wafer consists of 7 test chips after standard clean room fabrication. Fabrication can be carried out either with crystallographic etching in KOH or deep Si plasma etching. The detail of the fabrication process sequence is outlined:

A 220 um thick doubly polished Si wafer with 75 mm in diameter is coated by LPCVD SiNₓ on both sides. The two SiNₓ surface are patterned with masks as shown above. The wafer is then etched from both sides in hot KOH. After 110um of the silicon has been etched from each side, the paddle is free from the surrounding frame expect at the left-hand end. Etching is continued until the beam section has desired (40um) thickness after which the wafer is rinsed and dried. The paddle has the full thickness of the wafer. Cleavage streets 75 um deep (defined by the width of the patterned line on one mask) are etched from one side. The individual chips may be cleaved from the wafer at this point as the test chip carrier. Then either individually blanket metallized thin film, or the whole wafer may be blanket metallized can be deposited with desired thickness and film materials for the following tests of interested.

The structure can be metallized from both sides. No processing is required after the metal has been deposited. If desired, the SiNₓ may be removed by reactive ion etching (RIE) before metal deposition. Later, a metallic conducting surface is required on both sides of paddle for stable deflection and capacitance measurements to be made. Figure 3 shows the KOH process schematic.

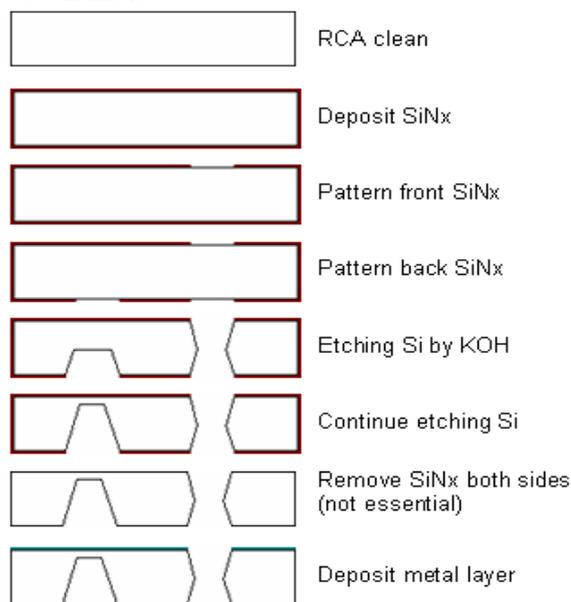

RCA clean

Deposit SiNx

Pattern front SiNx

Pattern back SiNx

Etching Si by KOH

Continue etching Si

Remove SiNx both sides (not essential)

Deposit metal layer

Figure 3   KOH etching process for an example of thin metal sample

Tapered sidewalls are naturally produced with KOH etching. The alternative is to use deep RIE. It produces the vertical sidewalls and the processes are clean and easy. Processing with deep Si etching is also a two-mask process. Etching from the front can define the paddle geometry through the full thickness of the wafer. Etching from the back creates the desired beam thickness. It may be desirable to the process with slightly non-vertical sidewalls to facilitate producing continuously conducting surfaces on both sides of a chip when it is metallized on tested thin film. Figure 4 shows the deep RIE process schematic.

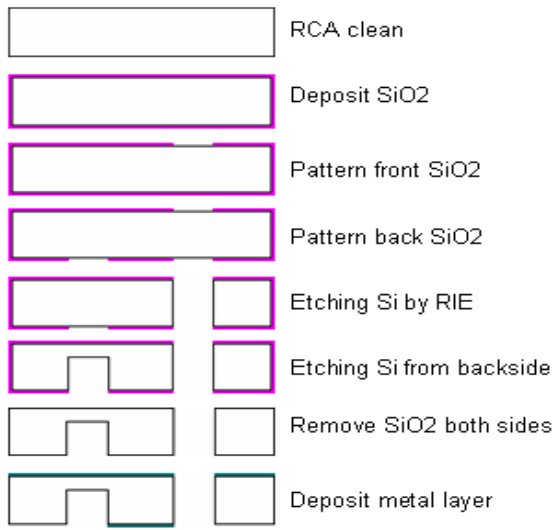

| | |
|---|---|
| | RCA clean |
| | Deposit SiO2 |
| | Pattern front SiO2 |
| | Pattern back SiO2 |
| | Etching Si by RIE |
| | Etching Si from backside |
| | Remove SiO2 both sides |
| | Deposit metal layer |

Figure 4: RIE etching process for an example of thin metal sample

The finish paddle sample is shown in Figure 5 and 6. The figure 5 is the whole sample view; the white zone is after KOH etching and the Figure 6 shows just paddle plate. The Figure 7 is the wafer just finish KOH etching.

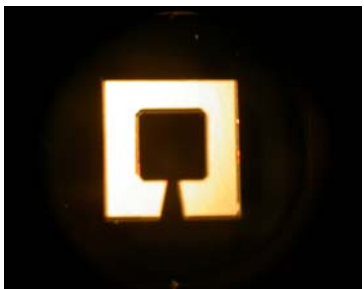

Figure 5: A finished paddle sample

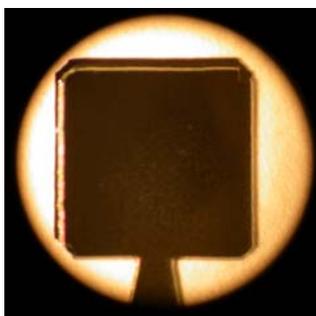

Figure 6: A finished paddle sample

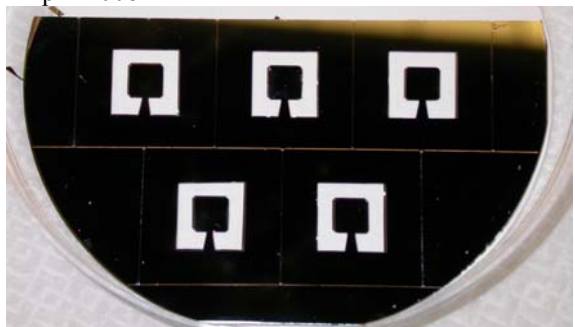

Figure 7: A finished paddle sample

### B. Testing System

The apparatus is custom design and the system set-up is design to measure beam deflection by capacitance. It consists of the guard-ringed capacitor electrode, the metal spacer, the calibration sample and the deflection electrode. Figure 8 shows the schematic of the measuring system. The sample chip is mounted together with a guard-ringed capacitor electrode as shown below. A spacing of 25 to 125um to the window frame chip surface around the paddle structure is defined by a metallic spacer.

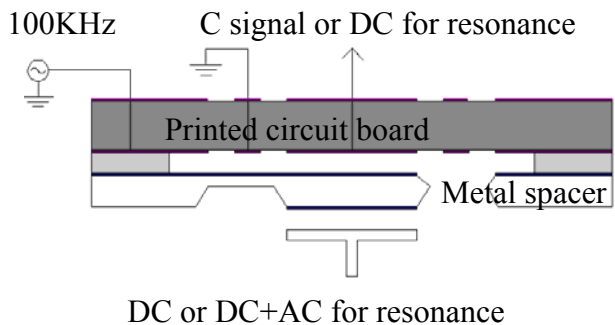

Figure 8: System schematic

During the experiment, we apply an electrostatic force to the specimen and measure the capacitance change. The deflection of the paddle beam can be measured from the capacitance value which is described by

$$C = \frac{\varepsilon_0 A}{d}$$

Where A is the capacitor plate area, d is the spacer thickness and $\varepsilon_0$ is dielectric constant. Thus we can have linear correlation between deflection space and the accurate capacitance value.

A second electrode is mounted below the paddle plate. This electrode is used for electrostatic deflection of paddle. The whole chip is at DC ground but driven at 100 kHz with amplitude of a few volts. That provides a displacement current to central electrode of the capacitor plate which is proportional to the capacitance, and hence inversely proportional to the gap. Depending on the spacing selected, the capacitance is between 2 and 4 pF. The measurement of the capacitance can be made to a precision of approximately 0.1 fF so that paddle spacing changes of 50 nm are readily determined. The paddle can be pulled up with a DC voltage on the guard-ringed electrode or pulled down with a DC voltage on the lower electrode. The capacitance measurement can be made with a time resolution of $\pm 10$ msec.

### C. Measurements of specimen bending and strain

For the electronic setup of the capacity measurement, a sine-wave generator at 100 kHz is applied to the film simultaneously measuring capacity of the paddle capacitor while a second generator drives at the same frequency for a test capacitor which has a known capacity. The two units are coupled (one master, one slave) and have a phase shift of $180^0$. Figure 9 shows the circuits. The $180^0$ out of phase currents from the two generator-capacity pairs are summed at input of change sensitive preamplifier. The amplified sum is measured with lock-in amplifier with the reference signal from one of the frequency generators.

When the current flowing through the two capacitors is approximately equal, the lock-in will show a value close to zero. In this case the ratio of the capacities is inversely-proportional to the ratio of the amplitudes of the driving:

$$V_1 \cdot C_1 \sim \frac{V_1}{Z_1} = \frac{V_2}{Z_2} \sim V_2 \cdot C_2$$

If now the capacitance of the paddle capacitor is changed by a change in $y_p$, the $180^0$ out of phase currents are unbalanced and the lock-in amplifier will measure the difference.

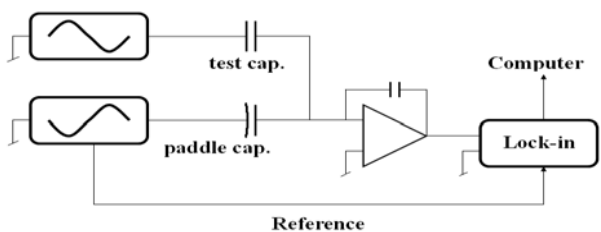

Figure 9: Electronic setup for the capacity measurement

The measurement is controlled by PC through National Instrument LabVIEW program. The control electronics include a controller, amplifier and waveform generator. Monitored signals are conditioned and then fed into an A/D board which is located in a PC. Data acquisition is performed with LabVIEW software. During sample testing, it is placed inside the chamber, and the samples are monitored with an optical microscope. After the sample is being locked inside the system, then wait until the system is reaching the thermal equilibrium and the capacitor read out is clear, the test can be perform.

### D. System Calibration

The calibration of the testing system is done to see if a linearity of capacitor value respect to the space between the paddle face and top electrode panel can be obtained. After the system is built, we can set up spacer of different thickness between 25 and 125um to hold the metal film on the sample at a fixed distance from metallic film on the Pyrex plate. Together with the metal film on the sample this plate forms a capacitor. A guard ring connected to ground surrounds the capacitor, shields it from stray signals and minimized fringe capacitance. The outermost region of the pattern metal film provides electrical connection to the sample. With different spacer we can calculate respected capacitance value at the same time we can also measuring the capacitance with different spacer and check the linear correlation between the capacitance value and the spacer thickness. Figure 10 shows an example of calculated capacitance value versus 1/d which can be used as the calibration reference line of the system.

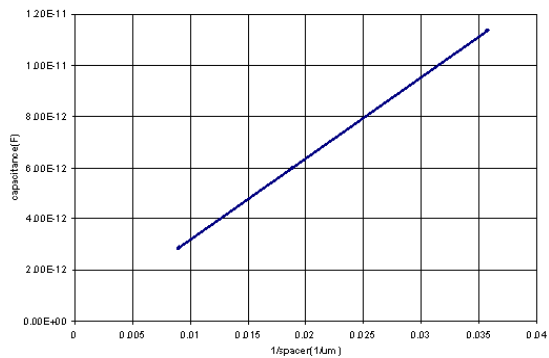

Figure 10 Calculated Capacitance versus 1/spacer thickness

## III. CALCULATION RESULTS & DISCUSSIONS

The mathematical studies of system performance were done to verify the performance of the systems. For all the measurement, the overall performance on the capacitance, electrostatic electrode and the paddle cantilever as shown in Figure 11 is essential.

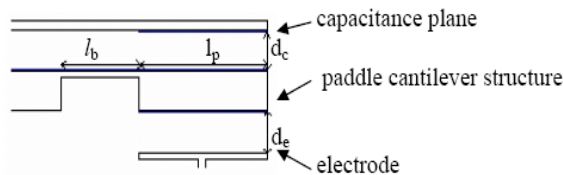

Figure 11 Schematic arrangements of capacitance, electrostatic electrode and the paddle cantilever

Whenever the cantilever is bent, the capacitance of the paddle cantilever structure is

$$C(y_b) = \varepsilon_0 l_p \int_0^{l_p} \frac{dx}{d_c - y_b - slope \cdot X}$$

where $\varepsilon_0$ is the dielectric constant of vacuum and $l_p$ is the width of the paddle. $y_b$ is the distance from the end position of the beam when it is flat to its position when the beam is bent the by a force (initial stress or electrostatic force). If the film has no initial stress $y_b=0$ and any electrostatic force will make it negative, If the film have initial tensile stress $y_b$ is positive with no electrostatic force and may be or negative with electrostatic deflection.

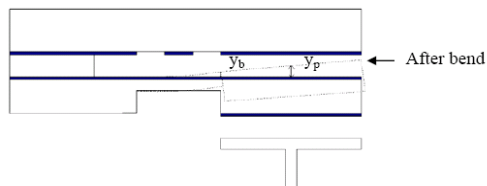

Figure 12 Schematic view of electrostatic force deflection.

In this case $y(X) = d_c - y_b - slope \cdot x$ and the slope of the paddle plane is $2y_b/l_b$. This yields

$$C(y_b) = \frac{\varepsilon_0 l_p l_b}{2 y_b} \left[ \ln \frac{d_e - y_b}{d_e - y_b - \frac{2 y_b l_p}{l_b}} \right]$$

$y_b$ and $y_p$ have the geometric relation

$$\frac{y_p}{y_b} = 1 + \frac{l_p}{l_b}$$

Thus we can use an example to calculate the sample deflection versus the capacitance. An example of the capacitance vs $y_p$ for a $\varepsilon_0$ of 8.85E-12 force/m, a $l_b$ of 3 mm, a $l_p$ of 5 mm, a $d_e$ of 100um. is shown in Figure 13.

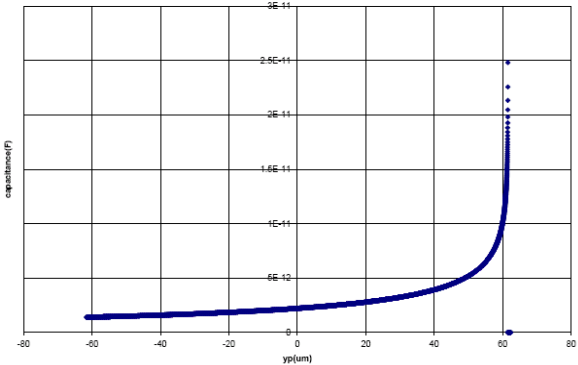

Figure 13: Capacitance versus $y_p$ position for $d_e = 100$um; $l_p = 5$mm; $l_b = 3$mm.

The electrostatic force on the paddle is given by

$$F_e = \frac{\varepsilon_0 l_p}{2} \int_0^{l_p} \frac{V^2 dx}{(d_e + y_b + slope \cdot X)^2}$$

The electrostatic force is always downward so that

$$\frac{F_e}{V^2} = \frac{\varepsilon_0 l_p l_b}{4 y_b} \left( \frac{1}{d_e + y_b} + \frac{1}{d_e + y_b + \frac{2 y_b l_p}{l_b}} \right)$$

Where $d_e$ is the distance from the paddle bottom plane to the electrode and V is the applied voltage. The electrostatic force divided by $V^2$ is shown in Figure 14 as a function of $y_b$.

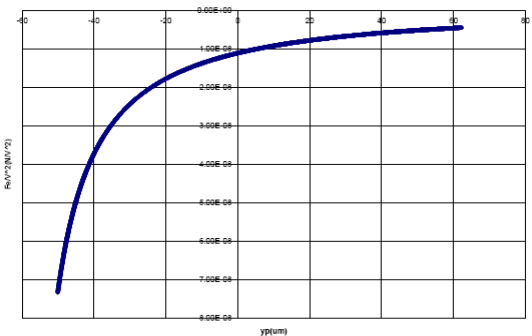

Figure 14: Electrostatic force versus $y_p$ position for $l_p = 5$mm; $l_b = 3$mm; $d_e = 100$um

For these dimensions the maximum value of $y_p$ is 61.53um because at that point the end of the paddle touches the capacitance plate. The minimum value is -61.53um because the paddle then touches the deflection electrode.

The total force on the paddle is

$$F_P^T = F_P^F + F_P^c - F_P^e$$

where the force on the paddle center due to stress in the film is

$$F_P^F = \frac{dE_F}{dy_p} = E_F \varepsilon_F V_F \frac{d\varepsilon_F}{dy_p}$$

and the compliance force of the beam

$$F_P^e = \frac{-1}{compliance} \cdot y_p , \quad compliance = \frac{y_p}{P} = \frac{6 l_b (l_b + l_p)}{E K t^3}$$

If we neglect electrostatic force on the paddle the total force is

$$F_P^T = F_P^F + F_P^C$$

We can write Eq. in the other word

$$F_P^F + F_P^C = E_F V_F \left( \varepsilon_F^0 + \varepsilon_b \right) K_1 - \frac{1}{compliance} y_p$$

Where $\varepsilon_F^0$ is the initial strain of the film depends on initial stress and $\varepsilon_b$ is a function of $y_p$. So we can written Eq. like below

$$F_P^F + F_P^C = E_F V_F \left( \varepsilon_F^0 - \frac{t_b y_p}{l_b (l_b + l_p)} \right) \frac{t_b}{l_b (l_b + l_p)} - \frac{E K t_b^3}{6 l_b (l_b + l_p)} y_p$$

Where $t_b$ is 40um; $l_b$ is 3mm; $l_p$ is 5mm; K is 0.3; E is biaxial young's modulus. The film and beam force together vs $y_p$ is shown in Figure 15.

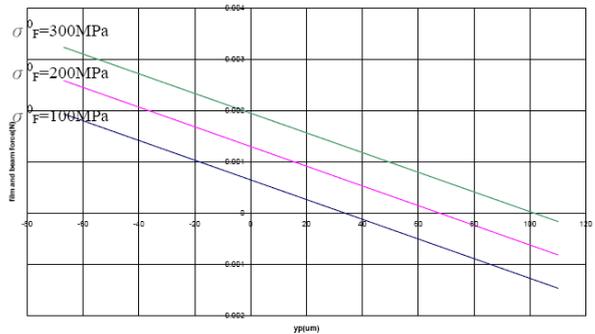

Figure 15: Film and beam force together versus $y_p$ position for initial stress in the film is 100, 200 and 300MPa

As we can observe in this plot when the force is equal to zero imply the system is in its equilibration position with no electrostatic force.

## IV. CONCLUSION

A novel "paddle" cantilever beam specimens with dimensions as few hundred nanometers to less than 10 nanometers is designed and fabricated. The experiment can be operated by using the electrostatic deflection on the "paddle" uniform distributed stress cantilever beam and then measure the deposited thin metal film materials on top of it. Capacitance techniques were used to measurement on the other side of the deflected plate to measure its deflection with respect to the force. The measured strain was converted through the capacitance measurement for the deflection of the cantilever. System performance on the residual stress measurement of thin films are calculated with three different forces on the "paddle" cantilever beam, including the force due to the film, compliance force and electrostatic force. The calculation helps to predict system performances including capacity versus bending high, driving voltage versus electrostatic force, maximum deflection and the free end of cantilever beam position in different residual stress. They also provided proof of the testing approach as well as potential use for the design and development of MEMS materials.

ACKNOWLEDGMENT

The authors are grateful to Prof. Walter Brown of Lehigh University for valuable discussions. This work was supported by Taiwan National Science Council; grant number NSC96-2221-E-005-082.